\documentclass[fleqn,usenatbib]{mnras}

\usepackage{newtxtext,newtxmath}
\usepackage{xcolor}
\usepackage{ulem}

\usepackage[T1]{fontenc}

\DeclareRobustCommand{\VAN}[3]{#2}
\let\VANthebibliography\thebibliography
\def\thebibliography{\DeclareRobustCommand{\VAN}[3]{##3}\VANthebibliography}

\usepackage{graphicx}	% Including figure files
\usepackage{amsmath}	% Advanced maths commands
\newcommand\hunits[1]{km s$^{-1}$ Mpc$^{-1}$}
\newcommand\kms[1]{km s$^{-1}$}
\newcommand\hmpc[1]{h$^{-1}$ Mpc}

\title[Impacts of bias in velocity reconstruction]{Biases in velocity reconstruction: investigating the effects on growth rate and expansion measurements in the local universe}

\author[Turner \& Blake]{
Ryan J. Turner$^{1}$\thanks{E-mail: rjturner@swin.edu.au} \&
Chris Blake$^{1}$
\\
$^{1}$Centre for Astrophysics and Supercomputing, Swinburne
University of Technology, Hawthorn, VIC 3122, Australia\\
}

\date{Accepted XXX. Received YYY; in original form ZZZ}

\pubyear{2023}

\begin{document}
\label{firstpage}
\pagerange{\pageref{firstpage}--\pageref{lastpage}}
\maketitle

\begin{abstract}
The local galaxy peculiar velocity field can be reconstructed from the surrounding distribution of large-scale structure and plays an important role in calibrating cosmic growth and expansion measurements. In this paper, we investigate the effect of the stochasticity of these velocity reconstructions on the statistical and systematic errors in cosmological inferences. By introducing a simple statistical model between the measured and theoretical velocities, whose terms we calibrate from linear theory, we derive the bias in the model velocity. We then use lognormal realisations to explore the potential impact of this bias when using a cosmic flow model to measure the growth rate of structure, and to sharpen expansion rate measurements from host galaxies for gravitational wave standard sirens with electromagnetic counterparts. Although our illustrative study does not contain fully realistic observational effects, we demonstrate that in some scenarios these corrections are significant and result in a measurable improvement in determinations of the Hubble constant compared to standard forecasts.
\end{abstract}


%%%%%%%%%%%%%%%%%%%%%%%%%%%%%%%%%%%%%%%%%%%%%%%%%%

%%%%%%%%%%%%%%%%% BODY OF PAPER %%%%%%%%%%%%%%%%%%

\section{Introduction}

The peculiar velocity field is an important ingredient in modern cosmological observations of the growth and expansion of the Universe, encoding information about gravitational physics, and adding noise to the smooth Hubble flow.  In both cases, reconstructions of the peculiar velocity field based on the observed large-scale structure play an important role, and therefore errors in these reconstructions can introduce both statistical and systematic errors to cosmological inferences.  In this paper we will consider a new general model for these error distributions, and present a first analysis of its impact on associated growth and expansion measurements.

Accurately measuring the redshift of the associated host galaxy of a distance indicator is a problem that all approaches to local measurements of the Hubble constant $H_0$ face, and is an unavoidable source of error \citep{Davis2019, Pesce2020}. The cosmological redshift of a galaxy will be distorted by its peculiar velocity and in the local Universe, where galaxy peculiar velocities are not negligible compared to the recession velocity due to the expansion of the Universe, is an important source of uncertainty in measurements of $H_0$. Recent work by \cite{Carr2022} and \cite{Peterson2022} has explored how improving redshift measurements, and more comprehensive modelling of the peculiar velocity component of redshift, can impact Type Ia supernova cosmology using Pantheon+ light curves \citep{Scolnic2022}.

The local peculiar velocity field is typically modelled from galaxy redshift survey data in order to predict the galaxy peculiar velocity \citep{Dekel1990, Willick1997, Zaroubi1999, Lavaux2016}, which is then subtracted from the measured host redshift to theoretically `correct' it and thus recover the actual cosmological redshift of the source \citep{Willick2001, Boruah2021}. Comparisons between the measured velocity field -- estimated from observations of the peculiar velocities of local galaxies -- and the predicted velocity and density fields -- inferred from galaxy redshift surveys -- also allow us to constrain the cosmological matter density parameter $\Omega_m$ and subsequently the growth rate of structure $f$ \citep{Strauss1995, Pike2005, Carrick2015, Said2020, Lilow2021, Courtois2023}. Given the importance of the modelled peculiar velocity in determinations of $H_0$ and $f$, we must also be conscious of any systematic or statistical error inherent to the creation of the model velocity field and how these errors may be propagated into cosmological measurements. \cite{Hollinger2021} investigate several facets of this approach, including choice of smoothing length and density tracer as well as the effects of noise. Close consideration of the linear theory commonly used to create the model velocity field from the galaxy density field reveals that the recovered local velocity field is a biased reconstruction of the measured velocity field. By accounting for this bias we can produce a model velocity field that matches observations, propagate this corrected model into our determinations of $H_0$, and quantify the effect this improvement has on measurement accuracy.

The current expansion rate of the Universe, parameterised by $H_0$, dictates key features of our Universe such as its size and age and is degenerate with parameters such as the dark energy equation of state, making $H_0$ itself a uniquely interesting cosmological parameter. There are two major approaches used to measure $H_0$, largely distinguished by the epoch of the Universe at which the measurement is made. Early-time measurements of $H_0$ are made by jointly fitting a set of cosmological parameters to large-scale features such as the cosmic microwave background \citep[CMB;][]{Planck2018Results}, baryon acoustic oscillations \citep[BAO;][]{ebosscollab2020}. Late-time measurements of $H_0$ are reliant on distance indicators such as Type IA supernovae (which act as standard candles) and a set of distance calibrations (i.e. parallax and Cepheid variables) that come together to form the local distance ladder. Other late-time measurements rely instead on the Tip of the Red Giant Branch (TRGB) as an important component of the distance ladder \citep{Freedman2019}.

Early-time predictions favour a smaller value of $H_0$, the most notable being the 2018 results of CMB experiment \textit{Planck} which find $H_0 = 67.27\,\pm\,0.60$ \hunits{} \citep{Planck2018Results}. Late-time predictions, however, favour larger values. One such recent measurement from the SH0ES team made with Hubble Space Telescope observations yields $H_0 = 73.04\,\pm\,1.04$ \hunits{} \citep{SH0ESHST2022}. A separate late-time measurement using the TRGB distance ladder finds $H_0 = 69.8 \pm 0.6$ (stat) $\pm 1.6$ (sys) \hunits{} \citep{Freedman2021}, a result within $2\sigma$ of SH0ES Cepheid calibrations and not statistically significant from CMB-inferred predictions. The discrepancy between measurements made at early- and late-times has led to introspection, and both approaches have been thoroughly probed for sources of systematic error; such as refining calibrations to the Cepheid distance ladder \citep{Pietrzynski2019, Reid2019, RiessLMC} and the TRGB distance ladder \citep{Anand2022, Anderson2023, Hoyt2023}, as well as modifications to physics at both early times \citep{Poulin2019, Kreisch2020, Jedmazik2021, Lin2021} and late times \citep{Zhao2017, Benevento2020} -- with no explanation fully resolving the disparity between the two approaches. We are left with a disagreement on the order of $4\sigma$ to $6\sigma$, depending on the data considered, that cannot be immediately explained away by systematic effects, termed the ``Hubble tension'' \citep[see][and references therein]{divalentino2021}.

Another alternative late-time approach to measuring $H_0$ is through the use of gravitational waves as standard sirens \citep{Schutz1986}. The distance to the source can be determined directly from the gravitational waveform, and the redshift may be measured from accompanying electromagnetic data. This method produces complementary results to those obtained using standard candles whilst not being dependent on the local distance ladder, and thus independent of any systematic errors inherent to standard candle measurements. Only one such event, GW170817, has had an electromagnetic counterpart detected with high confidence, and this was used by \cite{Abbott2017} to obtain the measurement $H_0 = 70.0^{+12.0}_{-8.0}$ \hunits{}. It was then shown by \cite{Howlett2020} that the choice of peculiar velocity for the host galaxy can significantly influence the resulting value of $H_0$, and combining a Bayesian model averaging analysis with constraints on the viewing angle obtained from radio data find that $H_0 = 64.8^{+7.3}_{-7.2}$ \hunits{}. An analysis undertaken by \cite{Mukherjee2021} applied statistical reconstruction techniques to obtain a peculiar velocity correction for the host galaxy of GW170817, and in combination with the same radio data find $H_0 = 68.3^{+4.6}_{-4.5}$ \hunits{}. Not only can the method used to account for the peculiar motion of galaxies influence the recovered value of $H_0$, especially when trying to obtain a result from a single source at low redshift, but it can also significantly impact the uncertainty in the estimate.

In Section \ref{sec:model_error} we introduce our new statistical model for the errors resulting from a linear-theory reconstruction of the velocity field.  In Section \ref{sec:growthfit} we explore the effect of these errors on determinations of the growth rate of structure based on this reconstructed velocity field, and in Section \ref{sec:indicators} we discuss how the velocity field reconstruction affects determinations of $H_0$, using a simulated sample of gravitational wave events with electromagnetic counterparts as distance indicators. We compare these simulated results to the expected error from analytical forecasts in Section \ref{sec:forecast}. Finally, we summarise our results and discuss future applications of this work in Section \ref{sec:conclude}.

\section{A model for the errors in velocity reconstruction}
\label{sec:model_error}

Velocity reconstruction is the process of inferring the peculiar velocity field across a cosmic volume from the estimated density field within that volume, typically mapped by galaxy positions.  In linear perturbation theory, neglecting redshift-space distortions, the $i = (x,y,z)$ components of the peculiar velocity field $v_i(\vec{x})$ are related to the matter overdensity field $\delta_m(\vec{x})$ via complex Fourier amplitudes:
\begin{equation}
  \tilde{v}_i(\vec{k}) = -iaHf \frac{k_i}{k^2} \tilde{\delta}_m(\vec{k}) ,
\label{eq:linear_theory}
\end{equation}
in terms of the cosmic scale factor $a$, Hubble parameter $H$ and growth rate $f$, where $k_i$ are the components of wavevector $\vec{k}$.  An example derivation of Eq. \ref{eq:linear_theory} can be found in Appendix C of \cite{2017MNRAS.471..839A}.  The overdensity field is typically constructed by gridding a density-field tracer and smoothing the resulting distribution to reduce noise.  If the tracer has a linear bias factor $b$, then the growth rate in Eq. \ref{eq:linear_theory} can be replaced by $\beta = f/b$. The accuracy of a linear-theory reconstruction model has been thoroughly investigated by many authors during the past two decades \citep[e.g.,][]{Zaroubi1999, 2001ApJ...549..688B, 2014JCAP...05..003O, Lavaux2016, Hollinger2021}.  We note that in this study we will neglect the effect of redshift-space distortions, which shift the observed positions of galaxies as a function of the growth rate.

\begin{figure}
  \includegraphics[width=\columnwidth]{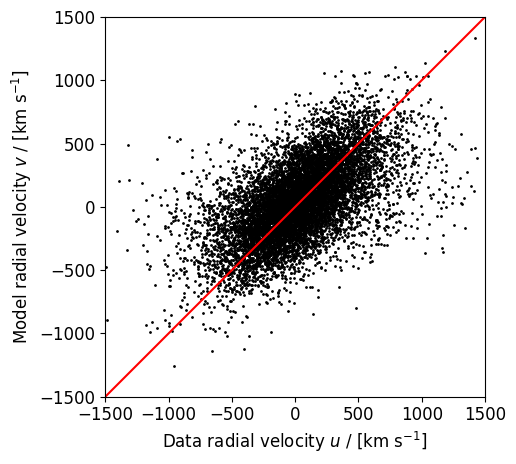}
  \caption{A comparison of the radial velocities $u$ of $10^4$ dark matter particles within a $1 \, h^{-3}$ Gpc$^3$ simulation box, and the model values $v$ at these positions applying linear-theory reconstruction to the density field in the box.  The diagonal solid line indicates $v = u$, about which the values show a significant scatter.}
  \label{fig:gigglez_velocity_comparison}
\end{figure}

In Fig. \ref{fig:gigglez_velocity_comparison} we show the result of applying Eq. \ref{eq:linear_theory} to reconstruct the model velocity field of a $1 \, h^{-3}$ Gpc$^3$ box of dark matter particles ($b = 1$) in today's Universe (snapshot with redshift $z=0$) created as part of the Gigaparsec WiggleZ N-body simulations \citep{2015MNRAS.449.1454P}, comparing the inferred velocity at each position with the underlying peculiar velocities of each particle in the simulation.  This simulation is chosen as an illustrative example, since we seek to describe generic trends.  In each case, we display the radial velocities of the particles relative to an observer.  For the purposes of this test, we reconstructed the density field using $10^5$ dark matter particles (where this number is chosen to create a non-negligible Poisson noise which we wish to model below) and sampled the velocity at the positions of $10^4$ different particles.  After binning the particles on a grid with dimension $128^3$, we applied a Gaussian smoothing to the density field with r.m.s. $\lambda = 10 \, h^{-1}$ Mpc as our fiducial choice, where we also compare results assuming $\lambda = 5$ and $20 \, h^{-1}$ Mpc.  After applying Eq. \ref{eq:linear_theory}, we imposed the appropriate complex conjugate properties on the velocity field. The result of the reconstruction is the characteristic distribution of Fig. \ref{fig:gigglez_velocity_comparison}, where the scatter is driven by breakdown of Eq. \ref{eq:linear_theory} (i.e.\ non-linear velocities) and Poisson noise in the density field.

\begin{figure}
  \includegraphics[width=\columnwidth]{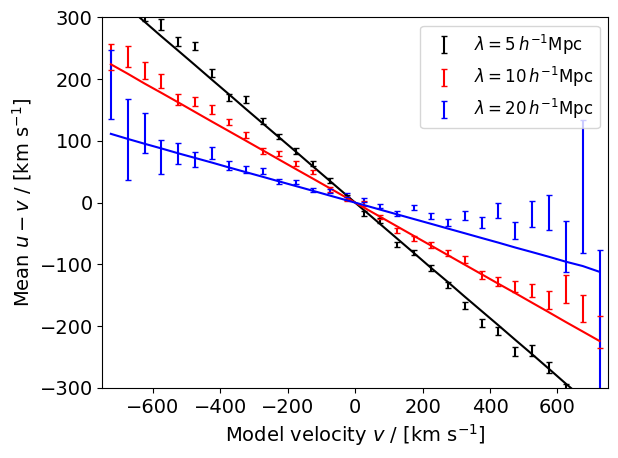}
  \caption{The mean difference between the radial velocity component of particles measured in the simulation and their value predicted by linear-theory reconstruction, binned as a function of model velocity.  Results are shown for three different smoothing scales $\lambda = (5, 10, 20) \, h^{-1}$ Mpc.}  The errors in each bin are calculated as errors in the mean, and the solid line is the prediction of Eq. \ref{eq:mean_u}.
  \label{fig:gigglez_meanveldiff}
\end{figure}

In this study we introduce a phenomenological statistical model to characterise the inaccuracies in the reconstructed velocity field.  We show that the coefficients of this model may be related to relevant physical effects and used to correct biases in the reconstructed velocity field, as a useful alternative to full non-linear modelling. Fig. \ref{fig:gigglez_velocity_comparison} motivates a 2D elliptical Gaussian statistical model for the reconstructed model velocity component $v$ and the measured velocity component $u$, where the joint probability can be expressed as:
\begin{equation}
  P(u,v) \propto \exp{\left[ - \frac{1}{2(1-r^2)} \left( \frac{u^2}{\sigma_u^2} - \frac{2 \, r \, u \, v}{\sigma_u \, \sigma_v} + \frac{v^2}{\sigma_v^2} \right) \right]} ,
\label{eq:u_vs_v}
\end{equation}
in terms of the variance of $v$, $\sigma_v^2 = \langle v^2 \rangle$, the variance of $u$, $\sigma_u^2 = \langle u^2 \rangle$, and the cross-correlation coefficient $r = \sigma_{uv}^2/(\sigma_u \, \sigma_v)$, where $\sigma_{uv}^2 = \langle u \, v \rangle$.  We assume that the variables have zero mean by isotropy, $\langle u \rangle = \langle v \rangle = 0$.  The statistical relation between the model velocity $v$ and measured velocity $u$ is hence characterised by the values of $\sigma_u$, $\sigma_v$ and $r$ at every point.  An immediate deduction from Eq. \ref{eq:u_vs_v} is that, for a point with model velocity $v$, the mean measured velocity $\overline{u} \ne {v}$.  In fact:
\begin{equation}
  \overline{u} = \int_{-\infty}^\infty P(u,v) \, du = \frac{r \, \sigma_u}{\sigma_v} \, v ,
\label{eq:mean_u}
\end{equation}
which can be considered as an unbiased model prediction at each point.  Fig. \ref{fig:gigglez_meanveldiff} verifies the prediction of Eq. \ref{eq:mean_u} by averaging the measured velocity of the $10^4$ particles displayed in Fig. \ref{fig:gigglez_velocity_comparison} in bins of the model velocity $v$.  The errors in each bin are calculated as errors in the mean, $\sigma/\sqrt{N}$, where $N$ is the number of particles in the bin and $\sigma$ is the standard deviation of the velocity differences.  We see that $\overline{u} \ne {v}$, and is well-modelled by Eq. \ref{eq:mean_u}, where the coefficients $\sigma_u$, $\sigma_v$ and $r$ are calculated as described below.

Linear theory gives us analytical expressions for the parameters of the model $\sigma_u$, $\sigma_v$ and $\sigma_{uv}$:
\begin{equation}
  \sigma_u^2 = \frac{a^2 H^2 f^2}{6V} \sum_{\vec{k}} \frac{P(\vec{k})}{k^2} ,
\label{eq:sigusq}
\end{equation}
\begin{equation}
  \sigma_v^2 = \frac{a^2 H^2 f^2}{6V} \sum_{\vec{k}} \frac{\left[ P(\vec{k}) + \frac{1}{n} \right] D^2(\vec{k})}{k^2} ,
\label{eq:sigvsq}
\end{equation}
\begin{equation}
  \sigma_{uv}^2 = \frac{a^2 H^2 f^2}{6V} \sum_{\vec{k}} \frac{P(\vec{k}) \, D(\vec{k})}{k^2} ,
\label{eq:siguv}
\end{equation}
where $V$ is the box volume, $P(\vec{k})$ is the model matter power spectrum at wavenumber $\vec{k}$, $D(\vec{k}) = e^{-k^2 \lambda^2/2}$ is the Fourier transform of the damping kernel used to smooth the density field before applying the linear-theory velocity reconstruction, and $n$ is the particle number density.  Eq. \ref{eq:sigusq}, the variance of the underlying field, is the standard expression for the velocity dispersion resulting from a given density power spectrum \citep{1988ApJ...332L...7G}.  Eq. \ref{eq:sigvsq}, the variance of the model velocity field, includes the additional effects of the shot noise contribution to the power ($1/n$) and the damping of power resulting from the smoothing ($D^2(\vec{k})$).  Eq. \ref{eq:siguv}, the cross-correlation between the two, includes just one power of the damping.

We evaluate Eq. \ref{eq:sigusq} to \ref{eq:siguv} as a sum over a grid of Fourier modes, not as an integral over $\vec{k}$-space ($\int \frac{d^3\vec{k}}{(2\pi)^3} \rightarrow \frac{1}{V} \sum_{\vec{k}}$) because the inverse powers of $k$ imply that the expressions are dominated by a small number of low-$\vec{k}$ modes.  We scale each mode by a factor representing the number of degrees of freedom it contains owing to the complex-conjugate properties of the density field.  Evaluating these results for the GiggleZ simulation using the fiducial matter power spectrum and cosmological parameters, we find for our fiducial smoothing $\lambda = 10 \, h^{-1}$ Mpc: $\sigma_u = 375.7 \,$ \kms{}, $\sigma_v = 290.8 \,$ \kms{} and $\sigma_{uv} = 241.5 \,$ \kms{}, i.e. $r = 0.53$.  For the other smoothing scales we consider, these coefficients have the values $(\sigma_v, \sigma_{uv}, r) = (367.9, 268.0, 0.52)$ for $\lambda = 5 \, h^{-1}$ Mpc and $(\sigma_v, \sigma_{uv}, r) = (223.0, 205.1, 0.50)$ for $\lambda = 20 \, h^{-1}$ Mpc (where the value of $\sigma_u$ is unchanged). The prediction of the velocity bias using Eq.\ref{eq:mean_u} with these coefficients is displayed in Fig.\ref{fig:gigglez_meanveldiff}, successfully validating these calculations for each of our considered smoothing scales for this example N-body simulation.

\section{Application to fitting a growth rate using velocity reconstruction}
\label{sec:growthfit}

An example of the potential impact of the statistical bias in the reconstructed velocity field, as represented by Eq. \ref{eq:mean_u}, is an analysis which uses a comparison between the measured and model velocities to fit for the growth rate of structure.  In this section we explore this potential systematic error using a test case. We note that this test case is designed to illustrate the utility of our phenomenological statistical model as an analysis framework, rather than realistically representing any real data sample or analysis.

In order to create a more accurate test of the linear theory expressions for $\sigma_u$, $\sigma_v$ and $\sigma_{uv}$, we generated 400 lognormal realisations of the density field with the same properties as the $1 \, h^{-3}$ Gpc$^3$ GiggleZ simulation box.  We generated the density field of each lognormal realisation from the fiducial $z=0$ matter power spectrum of the GiggleZ simulation over a $1 \, h^{-3}$ Gpc$^3$ cube, and derived each velocity field by transforming the density modes using Eq. \ref{eq:linear_theory} (taking care to preserve the correct complex conjugate properties). We smoothed the density field using the same smoothing scales used in Sec. \ref{sec:model_error}, $\lambda = (5, 10, 20) \, h^{-1}$ Mpc, and sampled the radial velocities at $N = 10^5$ random positions within the cube.

\begin{figure*}
  \includegraphics[width=2\columnwidth]{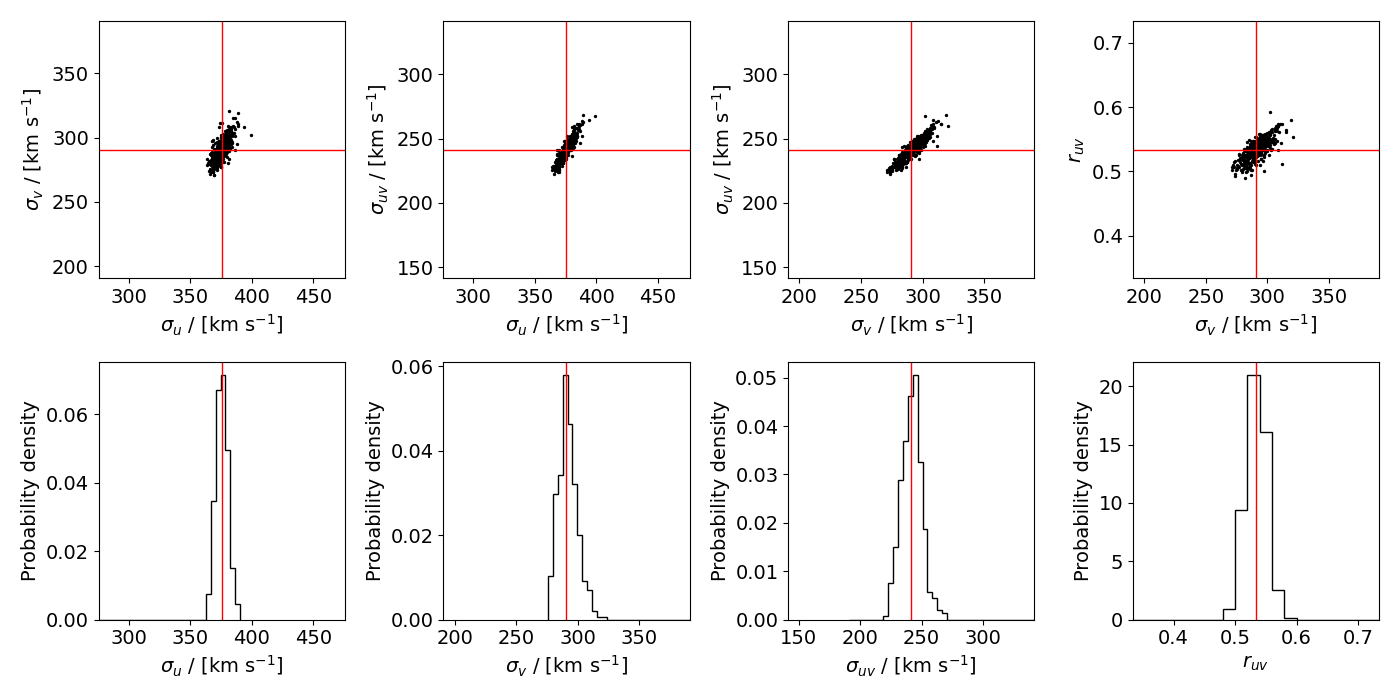}
  \caption{A comparison of the direct measurement of the coefficients of our statistical model relating the reconstructed and measured velocities, $\sigma_u$, $\sigma_v$ and $\sigma_{uv}$ (from which we deduce $r = \sigma_{uv}^2/(\sigma_u \, \sigma_v)$, with the analytical expressions of Eq. \ref{eq:sigusq} to \ref{eq:siguv}.  We measure the coefficients by sampling 400 lognormal realisations, and the panels of the figure display their joint distribution (top row), and a histogram (bottom row).  The solid red lines indicate the predictions of the analytical model, which are in good agreement with the measurements. These results use our fiducial smoothing scale, $\lambda = 10 \, h^{-1}$ Mpc.}
  \label{fig:lognormal_coefficients}
\end{figure*}

For each lognormal realisation we {\it estimate} the coefficients introduced in Sec. \ref{sec:model_error} as $\hat{\sigma}_u^2 = \overline{u^2}$, $\hat{\sigma}_v^2 = \overline{v^2}$ and $\hat{\sigma}_{uv}^2 = \overline{u \, v}$, where $u$ and $v$ are the measured and reconstructed radial velocities relative to the observer.  In Fig.  \ref{fig:lognormal_coefficients} we compare the distribution of estimated coefficients across the ensemble of simulations to the linear-theory calculations quoted in Sec. \ref{sec:model_error} for smoothing scale $\lambda = 10 \, h^{-1}$ Mpc, validating the accuracy of the model.

\begin{figure}
  \includegraphics[width=\columnwidth]{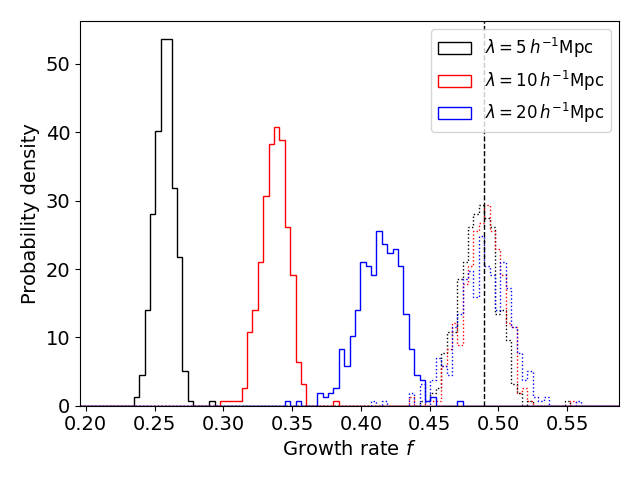}
  \caption{Histograms of best-fitting growth rates to the 400 lognormal realisations, using a velocity-density comparison method where the model velocity at each point is compared with the measured velocity in the simulation.  Results are shown for three different smoothing scales $\lambda = (5, 10, 20) \, h^{-1}$ Mpc.  The solid histograms are constructed assuming the model velocity is equal to the reconstructed velocity, and the dotted histograms are constructed after scaling the model velocities in accordance with Eq. \ref{eq:mean_u}.  The vertical black dashed line is the fiducial value of the growth rate.}
  \label{fig:lognormal_growthfits}
\end{figure}

We then fit the growth parameter $f$ to each lognormal dataset by minimising the chi-squared statistic,
\begin{equation}
    \chi^2 = \sum_i \left( \frac{u_i - (f/f_{\rm fid}) \, v_{{\rm mod},i}}{\sigma_i} \right)^2 ,
\end{equation}
where $i$ denotes the positions in the volume, and the velocity errors at each point are given by $\sigma_i = \sigma_u \sqrt{1-r^2}$.  We compare two options: specifying the model velocity at each location using the result of the linear-theory reconstruction, $v_{\rm mod} = v$, and scaling the reconstruction in accordance with Eq. \ref{eq:mean_u}, $v_{\rm mod} = \left( \frac{r \, \sigma_u}{\sigma_v} \right) v$.  Since the model velocities are proportional to the growth rate according to Eq. \ref{eq:linear_theory}, we can fit for an effective growth rate in each case by simply shifting the model by $f/f_{\rm fid}$, where $f_{\rm fid} = 0.49$ is the fiducial growth rate used to convert the density field to a velocity field for the lognormal realisation.  Fig. \ref{fig:lognormal_growthfits} displays the best-fitting growth rate according to each method and smoothing scale as a histogram over the lognormal realisations, indicating that the ``corrected'' velocity recovers an unbiased growth rate in all cases, whereas using the reconstructed velocity alone does not.

In Appendix \ref{sec:selection_function} we detail how this model can be modified in the presence of a survey selection function which modulates the tracer number density field.  In future work we will consider the presence of additional realistic observational effects such as this, together with other factors such as velocity measurement noise and redshift-space distortions.

\section{Application to distance indicators}
\label{sec:indicators}
Statistical bias in the reconstructed velocity field will also impact calculations of host galaxy redshifts that are needed to measure $H_0$. The local observed redshift ($z_{\rm obs}$) of a galaxy used to determine $H_0$ is the product of a cosmic expansion component ($z_{\rm cos}$) and a contribution from the galaxy peculiar velocity ($v_{\rm pec}$) in the form 
\begin{equation}
    (1 + z_{\rm obs}) = (1 + z_{\rm cos})(1 + v_{\rm pec}/c)\,,
\end{equation}
where $c$ is the speed of light. 

A single gravitational wave event will have the greatest constraining power at low redshift, else the error in $H_0$ will be dominated by the distance error \citep{Chen2018, Nicolaou2020}. For nearby sources, however, the impact of peculiar velocities on the observed redshift is more pronounced than for sources at high redshift, where the recessional velocity is much larger by comparison. Uncertainties in the peculiar velocity component can then negatively impact estimates of $H_0$, and if our reconstruction of the velocity field contains statistical or systematic errors we will propagate these errors into our estimates.

Measurements of distance are typically obtained from distance indicators such as standard candles, and an estimate of redshift obtained from accompanying electromagnetic data. Gravitational waves offer an alternative approach to estimating cosmic distances that, unlike standard candles, is not dependent on a local distance ladder. Binary inspiral models can be fit to gravitational wave signals in order to deduce the physical parameters of the event that produced the wave, and the amplitude of the waveform may be used as a 'standard siren' to estimate distance. If there is an electromagnetic counterpart associated with the compact binary coalescence then we may also infer a recessional velocity, and thus directly probe $H_0$ with only a single event \citep{Abbott2017, Hotokezaka2019, Nicolaou2020}. Gravitational waves do not strictly require a electromagnetic counterpart to be used to estimate $H_0$, and these `dark sirens' can be correlated with galaxy redshift catalogues to localise the signal and predict a host galaxy redshift \citep{Fishbach2019, Soares-santos2019, Finke2021, Mukherjee2021_ds, Palmese2021, Mukherjee2022}.

\cite{Mortlock2019} present a framework for determining $H_0$ from a sample of gravitational wave events with electromagnetic counterparts, which we rewrite as
\begin{equation}
    \hat{H}_0 = \frac{c \, z_{\rm cos} + x}{D}\,,
    \label{eq:H0_est}
\end{equation}
where the distance to the event is given by $D$, the cosmological redshift by $z_{\rm cos}$ and the velocity noise by $x$, which can represent different combinations of observational and model contributions as we define below.

The uncertainty in the distance measurement, $\sigma_D$, can be written as 
\begin{equation}
    \sigma_D \approx \frac{D^2}{D_*^2}\sigma_*\,,
    \label{eq:err_in_D}
\end{equation}
where $D_*$ is the maximum distance out to which gravitational wave sources can be measured by a survey, given some SNR threshold, and $\sigma_* = 2D_*/$SNR is the uncertainty in the distance measurement at $D = D_*$.

We can quantify the error in $H_0$ that we would expect to see as a result of using a biased reconstruction of the local peculiar velocity field, such as that obtained from Eq. \ref{eq:linear_theory} and \ref{eq:u_vs_v}, and compare this with the error in $H_0$ determinations obtained with a corrected velocity field reconstruction, via Eq. \ref{eq:mean_u}. We employ the same set of 400 lognormal realisations as before, each now reduced to a sphere with radius $150$ \hmpc{} centred on an observer and containing on average $1400$ objects. Each object in these realisations is associated with three values: the radial distance from the observer -- $D$ -- in units of \hmpc{}, the underlying peculiar velocity at the object's position -- $u$ -- in units of \kms{}, and the model velocity at that position obtained using the linear-theory reconstruction -- $v$ -- in units of \kms{}.

We first convert $v$ into an unbiased prediction of $u$ following Eq. \ref{eq:mean_u}, $\left(
\frac{\sigma_{uv}^2}{\sigma_v^2} \right) v$, using the previously derived values of $\sigma_u$, $\sigma_v$ and $\sigma_{uv}$. We further reduce the size of each spherical realisation with a distance cut, removing any objects that are within 5 \hmpc{} of the observer or are beyond some maximum distance set by $D_*$. The cut at the lower bound is made to prevent negative values of $H_0$ upon the addition of randomly sampled observational errors. From this cut catalogue we then subsample $N$ objects for our ensemble, where $N$ is the number of GW events in our pseudo-catalogue. We calculate the error in the distance for each of these $N$ objects individually by sampling from a Normal distribution where $\sigma_D$ is given by Eq. \ref{eq:err_in_D}. Assuming a fiducial value of $H_0 = 70.0$ \hunits{}, we can also estimate the value of $z_{\rm cos}$ for each object by inverting the Hubble-Lemaitre law, and produce redshift errors for all N objects by sampling from a Normal distribution where $\sigma_z = 10^{-4}$. We can then produce a noisy estimate of $z_{\rm cos}$ by adding this redshift error term to the previously estimated value. If the number of objects in the cut catalogue is less than the value of $N$ that we require as a minimum, however, then we do not perform this analysis on that catalogue and move on to the next realisation.
 
We find the best fitting value of $H_0$ by performing a modified least squares regression in both the $D$ and $z$ directions, i.e. by minimising a $\chi^2$ function of the form
\begin{equation}
    \chi^2(H_0) = \sum^N_{i=1}\frac{(cz_{\rm cos\,,i} + x - H_0D_i)^2}{c^2\sigma_{z,i}^2 + H_0^2\sigma_{D,i}^2 + \sigma_x^2}\,,
    \label{eq:chisq_minimise}
\end{equation}
where $z_{\rm cos\,,i}$ and $D_i$ are the noisy estimates of the cosmological redshift and distance to the i'th object in the sample respectively, $\sigma_{z,i}$ and $\sigma_{D,i}$ are their corresponding errors, $x$ is the velocity noise given in Eq. \ref{eq:H0_est} and $\sigma_x$ is the error in this velocity component (see Sec. \ref{sec:forecast}).

\begin{figure}
    \centering
    \includegraphics[width=\columnwidth]{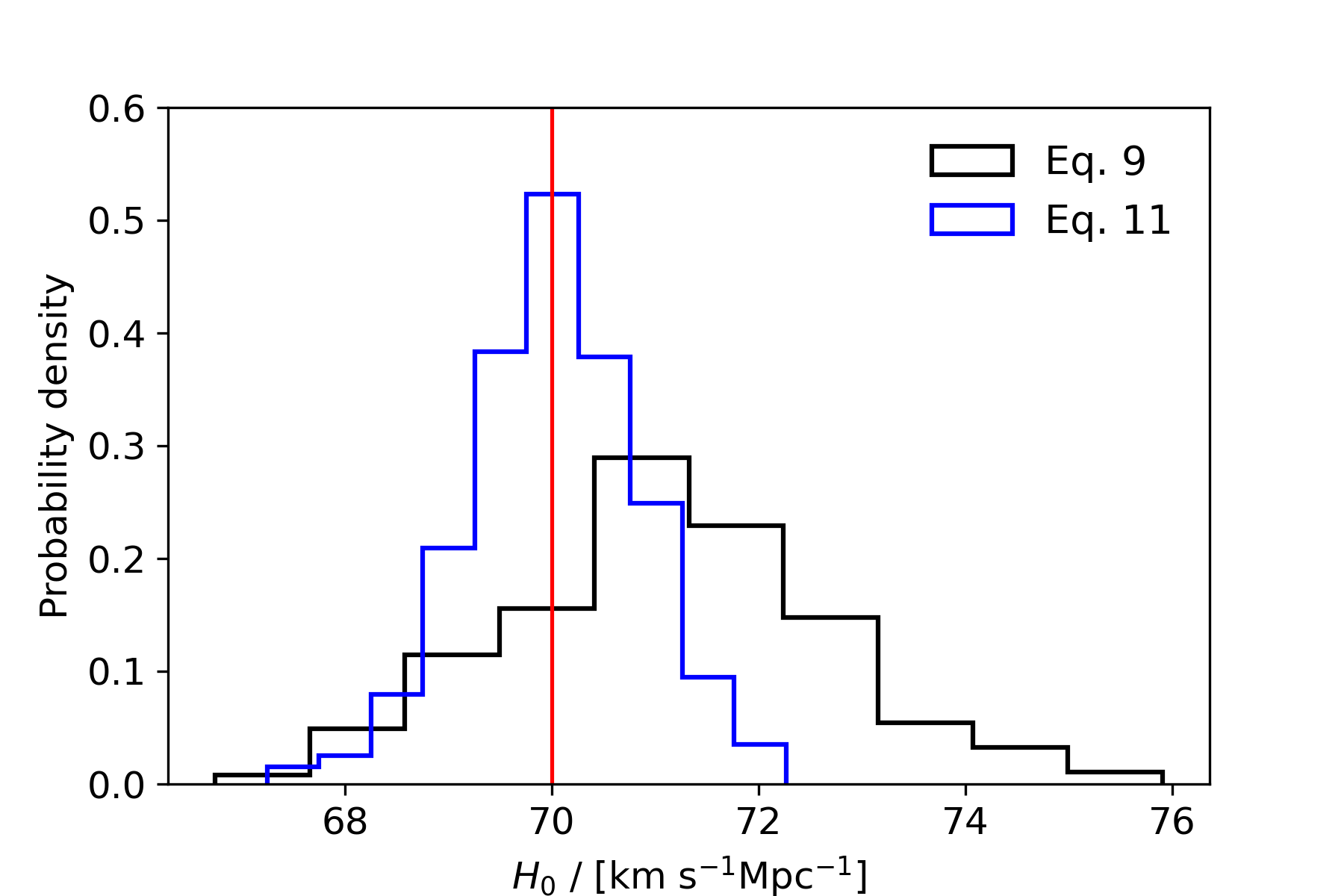}
    \caption{The distribution of the mean values of $H_0$ measured in each of the 400 lognormal mocks, setting $D_* = 150$ \hmpc{}. The values obtained from the `ratio' estimator Eq. \ref{eq:H0_est} are shown in blue, and those obtained using the maximum-likelihood estimator Eq. \ref{eq:chisq_minimise} are shown in black. The red, vertical line represents the fiducial value of $H_0 = 70$ \hunits{}.}
    \label{fig:eq8-10_hist}
\end{figure}

Adopting a maximum-likelihood estimator such as this, rather than explicitly using Eq. \ref{eq:H0_est} as the estimator for $H_0$, means that we avoid issues caused by dividing through by a noisy measurement such as the distance. Changes in the measurement of the distance caused by noise do not produce symmetric changes in value of $\hat{H}_0$. Calculations made with Eq. \ref{eq:H0_est} are therefore more sensitive to cases where noise reduces our measurement of $D$, resulting in an overestimate of $H_0$. We show the difference between these approaches in Fig. \ref{fig:eq8-10_hist}, using both equations to measure the mean value of $H_0$ from the same subsample of objects in each lognormal realisation. The maximum-likelihood approach produces results which are centred around the fiducial value of $H_0 = 70$ \hunits{} and are tightly distributed, and we find an average value of $H_0 = 70.02 \pm 0.042$ \hunits{}. In contrast, results from the `ratio' estimator are more sparsely distributed and are biased towards larger values of $H_0$ than the fiducial value, and we find an average value of $H_0 = 71.13 \pm 0.078$ \hunits{}. In this example we have defined the error to be $\sigma/\sqrt{N}$, where $N = 400$ corresponds to the number of lognormal realisations used to produce the $H_0$ measurements. The maximum-likelihood approach also requires the intercept of the fit to pass through the origin, which is an extra constraint not present in the simple estimator approach, driving the improvement in the error.

\begin{figure*}
    \centering
    \includegraphics[width=0.9\textwidth]{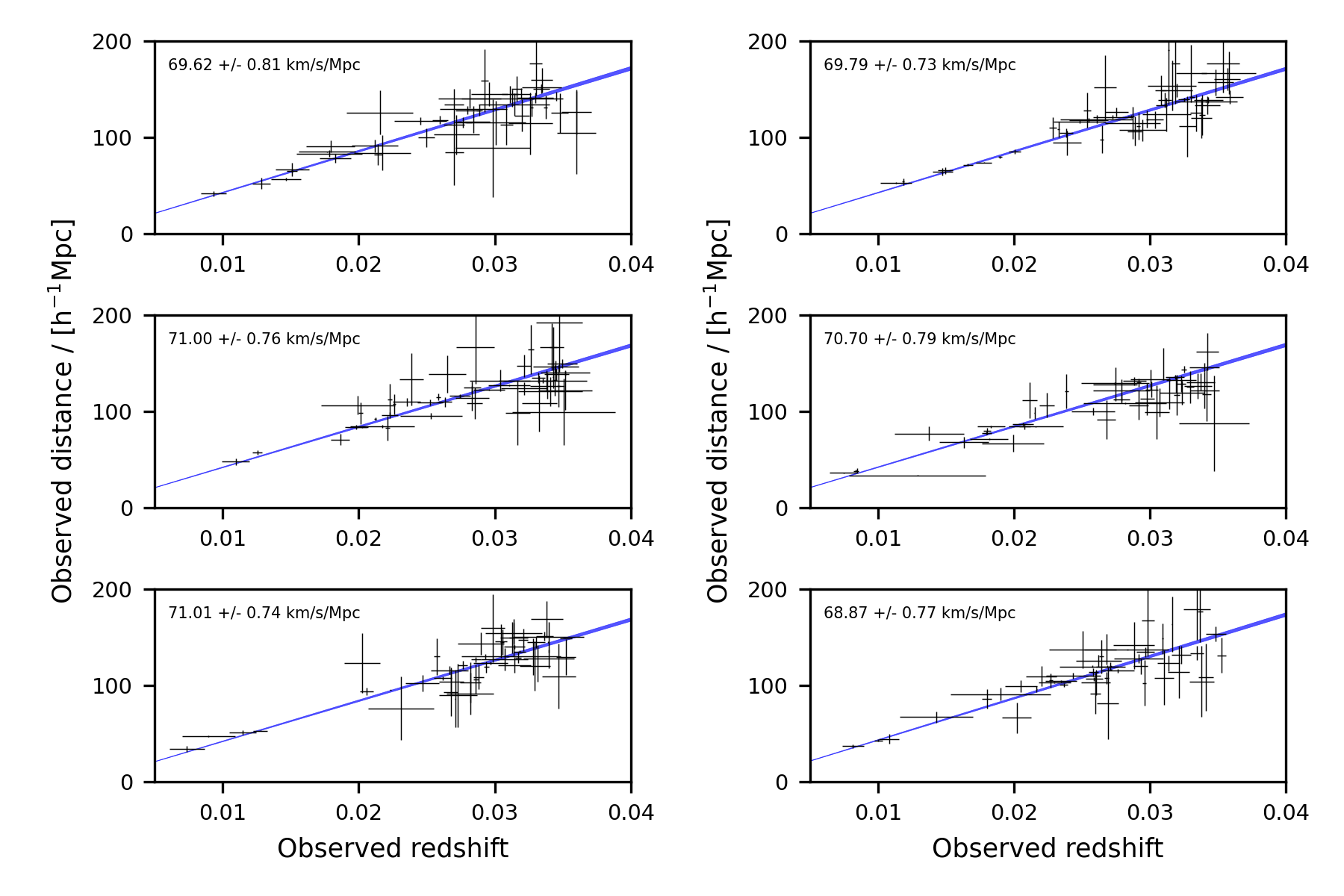}
    \caption{Determinations of $H_0$ obtained from Eq. \ref{eq:chisq_minimise} for six lognormal realisations selected randomly from our total ensemble of 400. These measurements are made at $D_* = 150$ \hmpc{} using $N = 50$ individual objects. The black crosses represent the error in the redshift (on x-axis) and distance (on y-axis) for each of the N objects, and are centred on their measured values ($z$, $D$). The blue region represents the 'measured' Hubble-Lemaitre law that uses the best-fitting value of $H_0$, and encompasses the $68\%$ confidence interval as defined by the standard deviation. The best-fitting value of $H_0$ and the standard deviation in the measurement obtained from each mock is also given in the top-left corner of their respective panels.}
    \label{fig:indiv_h0_lognormal}
\end{figure*}

Performing this fit for each realisation, and then taking the mean and standard deviation of the best-fitting values of $H_0$, we verify that we recover $H_0$ in an unbiased manner and also obtain an estimate of the error. We show this process for six randomly selected realisations in Fig. \ref{fig:indiv_h0_lognormal}, where we subsample each cut catalogue to select $N = 50$ objects from which we derive a measurement of $H_0$. 

We can compare different reconstructions of the model peculiar velocity field by changing how we consider the velocity noise $x$, and thus measure how these changes impact the error in estimations of $H_0$. There are three instances of $x$ that we are interested in, specifically. Setting $x_1 = u$ is the equivalent of adding a peculiar velocity component to $z_{\rm cos}$ and performing no model velocity correction, thus obtaining $z_{\rm obs} = z_{\rm cos}$. We can perform a linear-theory model velocity correction by setting $x_2 = u - v$. Finally, we can perform the model correction using the corrected velocity model by setting $x_3 = u - \left(
\frac{\sigma_{uv}^2}{\sigma_v^2} \right) v$. If the linear-theory model velocity reconstruction is a perfect reflection of the actual local velocity field, then we should recover $cz_{\rm cos}$ in case $x_2$. We know this is not the case from Fig. \ref{fig:gigglez_meanveldiff}, and so instead expect to recover $cz_{\rm cos}$ from the corrected model velocity case $x_3$.

\begin{figure}
  \includegraphics[width=\columnwidth]{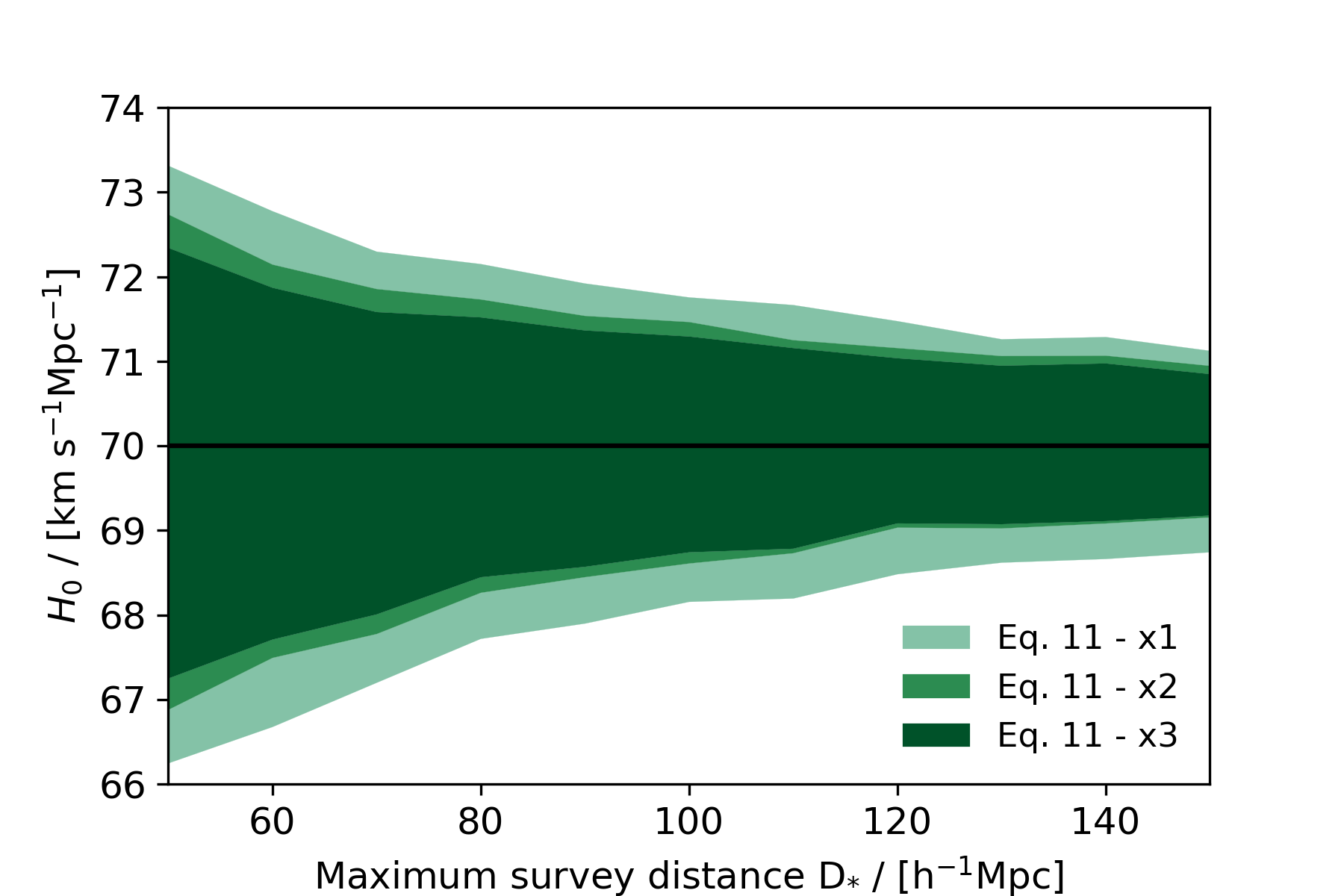}
  \caption{Measurements of $H_0$ made as a function of survey distance $D_*$, obtained by minimising the $\chi^2$ function described by Eq. \ref{eq:chisq_minimise}. The horizontal black line depicts the fiducial value of $H_0 = 70$ \hunits{}, while the coloured lines represent the mean values of $H_0$ at each value of $D_*$ for each of the three considered model correction cases. The shaded regions depict the standard deviation across the ensemble of $H_0$ measurements about the mean, made at each $D_*$. The colour representing each case is given by the legend in the bottom-right corner.}
  \label{fig:h0_fits}
\end{figure}

\begin{figure}
  \includegraphics[width=\columnwidth]{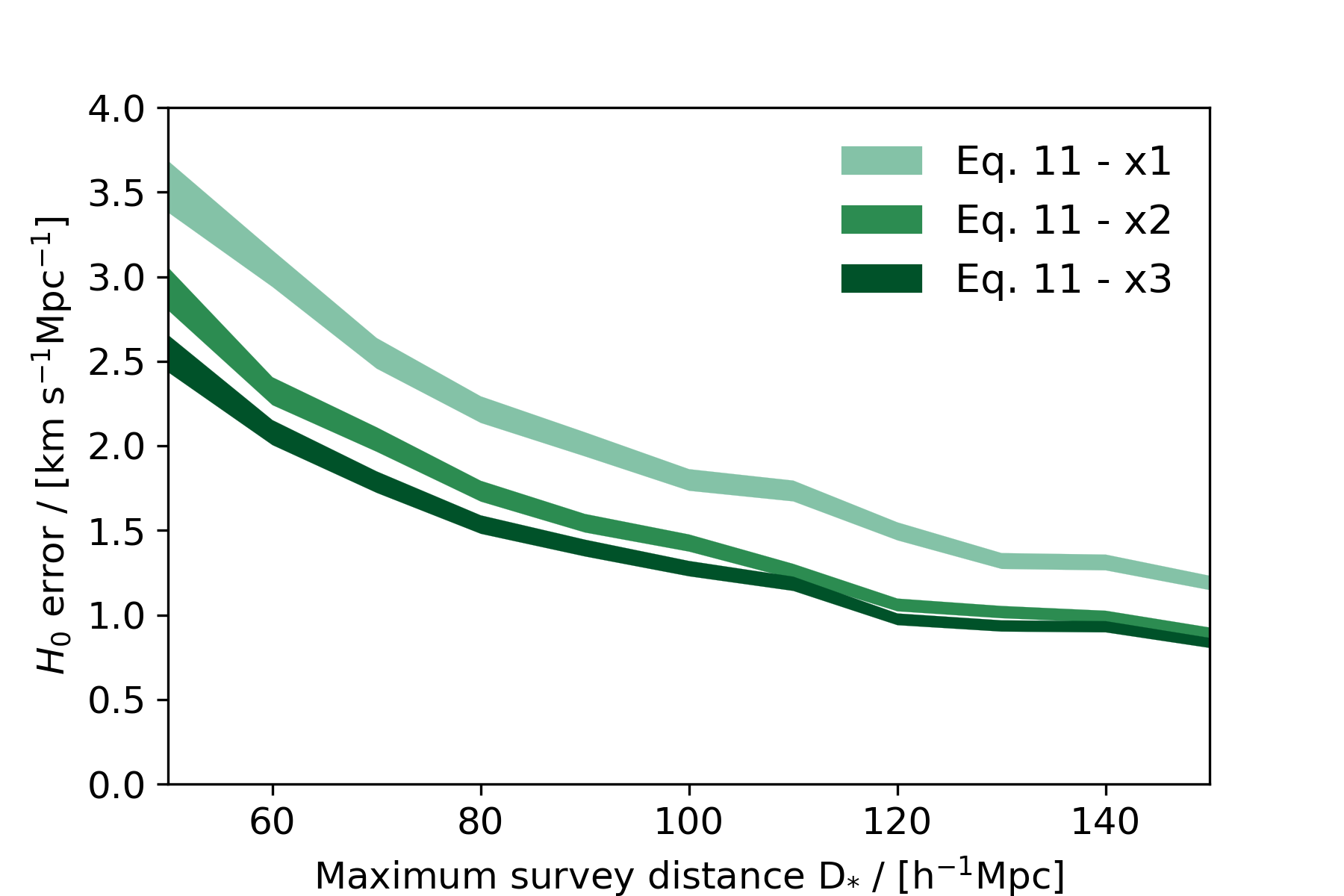}
  \caption{The standard deviation across the set of $H_0$ measurements as a function of $D_*$, the same values are also given in Fig. \ref{fig:h0_fits}. The shaded region represents the analytical error in the error given by the formula $\delta\sigma = \sigma/\sqrt{2(N-1)}$, where N is the number of $H_0$ measurements made in each $D_*$ bin. The colour of each region matches that in Fig. \ref{fig:h0_fits}.}
  \label{fig:h0_scatter}
\end{figure}

By repeating this process for a range of $D_*$ values -- from 50 \hmpc{} to 150 \hmpc{} in steps of 10 \hmpc{} -- and for all three cases of $x$ described above while keeping $N$ fixed, we can determine how the error in $H_0$ changes with the different model velocity corrections, and as a function of distance. The mean values of $H_0$ for each case, measured using Eq. \ref{eq:chisq_minimise}, are shown in Fig. \ref{fig:h0_fits}. The shaded regions in Fig. \ref{fig:h0_fits} represent the standard deviation across the $H_0$ fits in each $D_*$ bin, and are additionally shown in Fig. \ref{fig:h0_scatter}. At the maximum survey distance $D_* = 150$ \hmpc{}, the error in $H_0$ that we measure for the case $x_1$ is $\sigma_{H,x_1} = 1.190$ \hunits{}, for the case $x_2$ is $\sigma_{H,x_2} = 0.896$ \hunits{}, and for the case $x_3$ is $\sigma_{H,x_3} = 0.836$ \hunits{}.

Given the similarity in the errors in Fig. \ref{fig:h0_fits}, we also present the `error in the error' in Fig. \ref{fig:h0_scatter} as shaded areas about the standard deviation. We calculate this additional error analytically using the equation $\delta\sigma = \sigma/\sqrt{2(N-1)}$, where $N$ is the number of $H_0$ results in each bin. 

\section{Forecasts}
\label{sec:forecast}

\begin{figure}
  \includegraphics[width=\columnwidth]{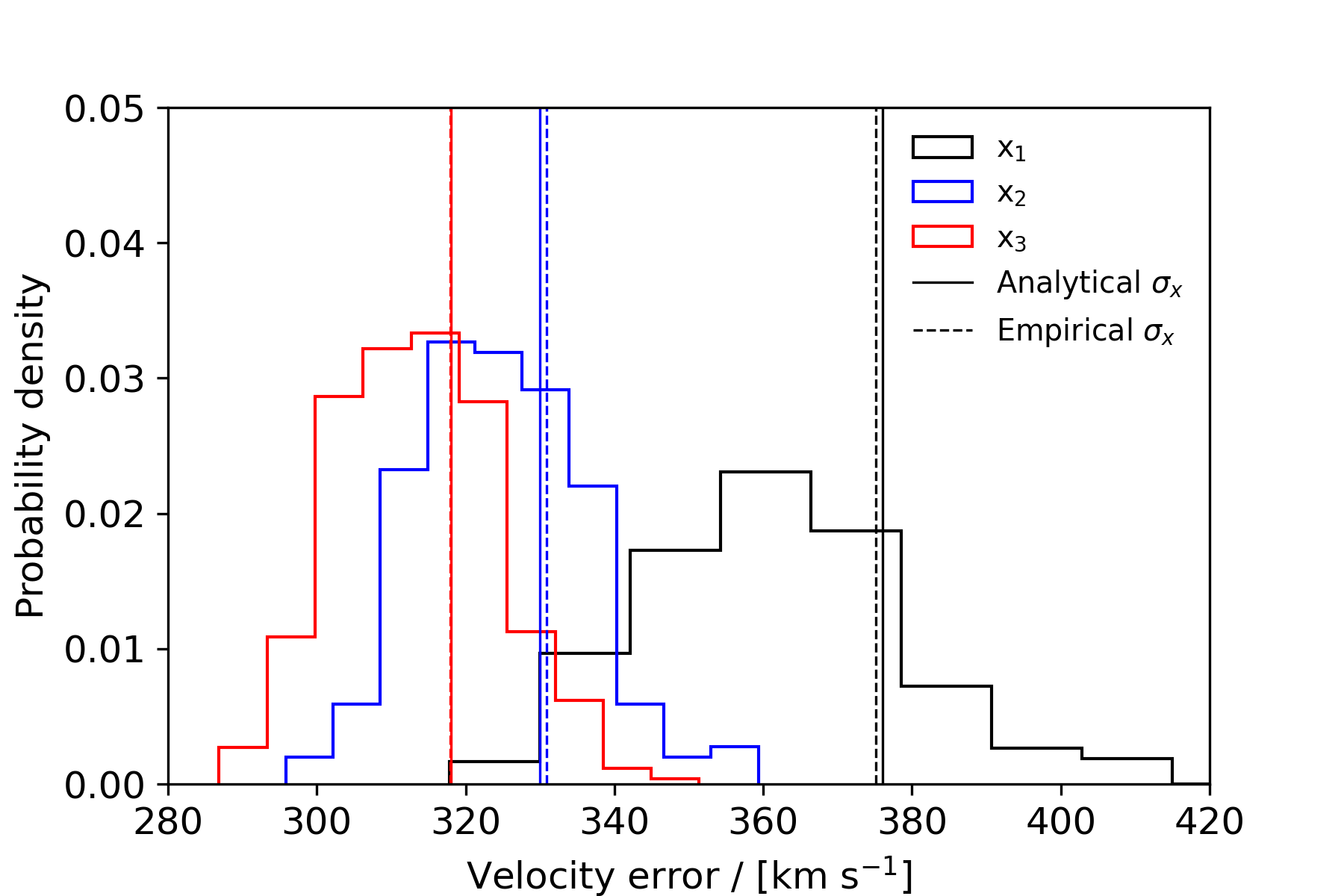}
  \caption{The distribution of velocity errors $\sigma_x$ in each treatment of $x$ discussed in Sec. \ref{sec:indicators}, calculated for each of the 400 lognormal realisations. The case $x_1$ is shown in black, $x_2$ in blue, and $x_3$ in red. The solid vertical lines represent the analytical error for each case, given in Sec. \ref{sec:forecast}, while the dashed vertical lines represent the numerically derived error when considering every object across all 400 realisations together.}
  \label{fig:sigmax_numeric}
\end{figure}

\cite{Mortlock2019} define the expected uncertainty in a measurement of $H_0$ from a sample of $N$ objects as
\begin{equation}
    \sigma_H = \frac{1}{N^{1/2}}\left(\frac{3}{5}\right)^{1/2}\frac{H_0\,\sigma_*}{D_*}\left(5\frac{D_0^2}{D_*^2} + 1\right)^{1/2}\,,
    \label{eq:sigH_mortlock}
\end{equation}
where $D_0$, the distance at which $\sigma_z$ and $\sigma_x$ become unimportant, is
\begin{equation}
    D_0 = \frac{\left(c^2\sigma_z^2 + \sigma_x^2\right)^{1/2}}{\left(\frac{H_0\,\sigma_*}{D_*}\right)}\,.
    \label{eq:d0_mortlock}
\end{equation}
We can therefore compare the measurement errors we obtain in Section \ref{sec:indicators} to the theoretically `ideal' errors that we should expect from Eq. \ref{eq:sigH_mortlock}. For this comparison we set $H_0 = 70$ \hunits{}, $D_* = 150$ \hmpc{}, $N = 50$, $\sigma_z = 10^{-4}$ and SNR $= 12$. The uncertainty in the velocity $\sigma_x$ will change depending on the form that our model subtraction $x$ takes, using the three different forms of $x$ discussed in Sec. \ref{sec:indicators},
\begin{enumerate}
    \item No model subtraction: $x_1 = u$\,,
    \item Model subtraction: $x_2 = u - v$\,,
    \item Corrected subtraction: $x_3 = u - \left(
    \frac{\sigma_{uv}^2}{\sigma_v^2} \right) v$\,.
\end{enumerate}
In the no model correction case $x = u$ and so we simply recover $cz_{\rm cos} + u = cz_{\rm obs}$, while in the two other scenarios we subtract a model velocity component from $u$ in an attempt to recover $cz_{\rm cos}$. We can analytically define the error in the velocity $\sigma_x$ for each correction scenario, 
\begin{enumerate}
    \item No model subtraction: $\sigma_{x_1}$ = $\sigma_u$ = 376 \kms{}\,,
    \item Model subtraction: $\sigma_{x_2}$ = $\sqrt{\sigma_u^2 - 2\sigma_{uv}^2 + \sigma_v^2}$ = 330 \kms{}\,,
    \item Corrected subtraction: $\sigma_{x_3}$ = $\sqrt{\sigma_u^2-\frac{\sigma_{uv}^4}{\sigma_v^2}}$ = 318 \kms{}\,,
\end{enumerate}
where $\sigma_u$, $\sigma_v$, $\sigma_{uv}$ are again as previously defined. Thus, we find that the error in the velocity noise $\sigma_x$ is minimised when subtracting our new corrected model from the observed redshift. We confirm these results empirically by measuring the standard deviation in the measurements of each definition of $x$ across every object in all 400 lognormal realisations, as well as plotting the distribution in the standard deviations recovered from each realisation individually too, see Fig. \ref{fig:sigmax_numeric}. The values of $\sigma_x$ we obtain numerically are all in agreement with the corresponding analytically derived values given above to within $< 1\%$. We can also show that the corrected model subtraction is in fact the optimal choice. Setting $x = u - \alpha v$, it follows that
\begin{equation}
    \sigma_x^2 = \sigma_u^2 - 2 \alpha\sigma_{uv}^2 + \alpha^2 \sigma_v^2.
\end{equation}
If we then vary $\alpha$ such that we minimise $\sigma_x$, we find
\begin{equation}
    \frac{d\sigma_x^2}{d\alpha} = 0 \rightarrow \alpha = 
    \frac{\sigma_{uv}^2}{\sigma_v^2} = r \,\frac{\sigma_u}{\sigma_v}\,,
\end{equation}
confirming the results we obtain in Eq. \ref{eq:mean_u} and Fig. \ref{fig:sigmax_numeric}.

Replacing the value of $\sigma_x$ in Eq. \ref{eq:d0_mortlock} with these three analytically derived quantities, we obtain $\sigma_{H,x_1} = 1.419$ \hunits{}, $\sigma_{H,x_2} = 1.388$ \hunits{}, and $\sigma_{H,x_3} = 1.380$ \hunits{}. We can compare these analytically derived errors to those we obtain empirically from the lognormal realisations, and from the ratio of these two values determine how well our approach in Sec. \ref{sec:indicators} works against the forecast measurement error. In the case of applying no model subtraction, $x_1$, the empirical error is $1.190/1.419 = 0.839 \approx 16\%$ smaller than the analytical error. Similarly, in the $x_2$ case the empirical error is $35\%$ smaller than the analytical error, and in the $x_3$ case the error is $39\%$ smaller. We summarise this comparison in Tab. \ref{tab:sigH_comparison}. 
\begin{table}
    \centering
    \begin{tabular}{c|c|c|c}
        $\sigma_x$ (\kms{}) & Forecast $\sigma_H$ & Measured $\sigma_H$ & Measured / Forecast\\
        \hline
        $\sigma_{x,1} = 376$ & 1.419 & 1.190 & 0.839\\
        $\sigma_{x,2} = 330$ & 1.388 & 0.896 & 0.646\\
        $\sigma_{x,3} = 318$ & 1.380 & 0.836 & 0.606
    \end{tabular}
    \caption{The $H_0$ error, in units of \hunits{}, obtained analytically and empirically using the three values of $\sigma_x$ derived in Sec. \ref{sec:forecast}. For each $\sigma_x$ we also compute the ratio between the analytical and empirical results.}
    \label{tab:sigH_comparison}
\end{table}

Rather than comparing these errors as analytical against empirical we can also perform a `meta-comparison', case against case, to see how these methods behave. Analytically, comparing the $x_3$ error to the $x_1$ error we see that $\sigma_{H,x_3}/\sigma_{H,x_1} = 1.380/1.419 \approx 0.973$ is approximately $2.5\%$ smaller, while empirically we find the error $\sigma_{H,x_3}/\sigma_{H,x_1} = 0.836/1.190 \approx 0.703$ is approximately $30\%$ smaller. The analytical error in $H_0$ obtained from Eq. \ref{eq:sigH_mortlock} is largely resistant to changes in the value of $\sigma_x$ we use, and thus is not sensitive to the choice of model subtraction we apply. The maximum-likelihood method, however, is sensitive to our choice of $\sigma_x$. The error in $H_0$ is much improved when applying any form of model subtraction, biased or otherwise, although the most accurate measurement of $H_0$ is made using the corrected model subtraction.

We are able to significantly improve the uncertainty in our predictions of $H_0$ by making some different choices in constructing our estimator, modifying our model subtraction so that it is optimal, and being mindful to limit the impact of distance errors.
\section{Conclusions}
\label{sec:conclude}

We have introduced a new statistical model for the reconstructed velocity field that minimises the velocity dispersion when removing the component of observed galaxy redshifts attributable to their peculiar motion. Applying this new model to mock cosmological tests; comparing the measured and model velocity fields to fit for the growth rate of structure and measuring the current rate of expansion from a sample of gravitational wave events, we have shown the potential impact that systematic error in the model velocity field can have when estimating cosmological parameters.

Standard candles and standard sirens captured by next-generation instruments will be used to obtain precision measurements of the Hubble constant in the local Universe. Errors present at low redshift will have a more significant impact on determinations of $H_0$ than the same error at high redshift. We believe it is prescient, then, to ensure that galaxy redshifts properly account for the impact of peculiar velocities on redshift measurements. Commonly applied linear-theory approximations of the velocity field are inherently biased, and do not replicate the mean measured velocity field as we would expect them to. Biases in our reconstructions of the velocity field propagate into cosmological results that are in some way informed by our understanding of the velocity field, most notably the growth rate of structure and the current expansion rate of the Universe. By correcting this discrepancy between the modelled and measured velocity fields, we can mitigate the error in our estimations of cosmological parameters due to galaxy peculiar velocities.

We have explored a framework in which we correct for this bias in the reconstructed velocity field and use the corrected field to estimate the value of $H_0$ for a sample of mock gravitational wave events. As part of this framework, we replace the `ratio' estimator with a maximum-likelihood estimator that jointly considers error in the redshift, distance and velocity observations. Measurements made with the `ratio' estimator are largely influenced by the error in the distance, and by alleviating this we produce a measurement of $H_0$ which is more accurate and unbiased with regard to our fiducial cosmology. 

We also compare the errors we obtain numerically with the error expected from analytical forecasts, specifically that set out by \cite{Mortlock2019} which is founded upon the `ratio' estimator Eq. \ref{eq:H0_est}. The errors we obtain are an improvement on this theoretical forecast, and at a maximum survey distance of $D_* = 150$ \hmpc{} we find that the uncertainty in our estimate of $H_0$ is approximately $39\%$ smaller than predicted by the forecast. In addition, we are able to prove analytically that the correction to the reconstructed velocity field we present is indeed the optimal correction to make, and we also verify this claim empirically.

In our testing of this framework we have neglected some forms of observational error such as velocity measurement noise and the effect of a survey selection function. Going forward it will necessary to apply this corrected velocity model to datasets with more realistic error considerations than our suite of realisations. We also believe there are implications for BAO reconstruction techniques, and would like to explore the parallels between such techniques and this work in the future.

\section*{Acknowledgements}

We thank the anonymous referee for constructive suggestions which improved the paper. RJT would like to acknowledge the financial support received through a Swinburne University Postgraduate Research Award and Australian Research Council Discovery Project DP220101610.

%%%%%%%%%%%%%%%%%%%%%%%%%%%%%%%%%%%%%%%%%%%%%%%%%%
\section*{Data Availability}

The data underlying this article will be shared on reasonable request to the corresponding author.


\bibliographystyle{mnras}
\bibliography{references}

%%%%%%%%%%%%%%%%% APPENDICES %%%%%%%%%%%%%%%%%%%%%

\appendix

\section{Including a selection function}
\label{sec:selection_function}

We now suppose that the density-field tracer we are using for the velocity-field reconstruction has a non-uniform distribution characterised by a selection function $W(\vec{x})$, which is normalised such that $\sum_{\vec{x}} W(\vec{x}) = N$, the total number of tracers.  We could infer the overdensity field as $\delta(\vec{x}) = N(\vec{x})/W(\vec{x}) - 1$, but this approach runs into difficulties when $W(\vec{x})$ is close to zero.  Therefore, analogously to power spectrum estimation \citep[e.g.,][]{1994ApJ...426...23F}, we instead adopt a density measure,
\begin{equation}
  \delta_N(\vec{x}) = \frac{N(\vec{x}) - W(\vec{x})}{N_0} ,
\label{eq:delta_withsel}
\end{equation}
where $N_0$ is the average of $N(\vec{x})$ across the survey volume, and re-construct the velocity field using $\delta_N(\vec{x})$ in Eq. \ref{eq:linear_theory}.  In the presence of a survey selection function, Eq. \ref{eq:sigvsq} is modified to:
\begin{equation}
  \sigma_v^2 = \frac{a^2 H^2 f^2}{6V} \sum_{\vec{k}} \left[ {\rm Conv} \left( \frac{1}{k^2} , |\tilde{W}(\vec{k}|^2 \right) P(\vec{k}) + \frac{1}{k^2} \, \frac{1}{n} \right] D^2(\vec{k}) ,
\label{eq:sigvsq_withsel}
\end{equation}
where ${\rm Conv} \left( \frac{1}{k^2} , |\tilde{W}(\vec{k})|^2 \right)$ denotes the convolution of $\frac{1}{k^2}$ and the square of the Fourier transform of the window function, $\tilde{W}(\vec{k})$.  This convolution is created by the use of Eq. \ref{eq:delta_withsel} to estimate the density field used in the reconstruction, analogously to the convolution of the model which is the expectation of a power spectrum estimate \citep{1994ApJ...426...23F}.

\begin{figure*}
  \includegraphics[width=2\columnwidth]{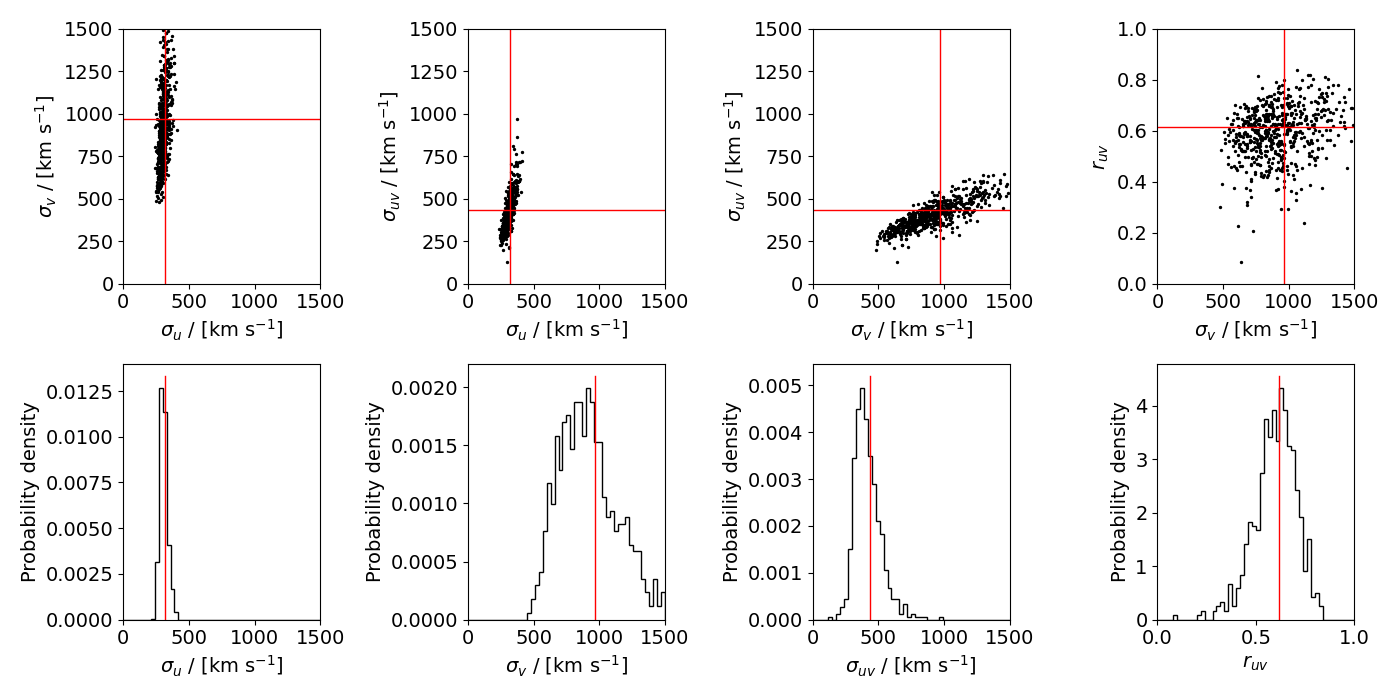}
  \caption{A comparison of the direct measurement of the coefficients of our statistical model relating the reconstructed and measured velocities, $\sigma_u$, $\sigma_v$ and $\sigma_{uv}$ (from which we deduce $r = \sigma_{uv}/(\sigma_u \, \sigma_v)$, with the analytical expressions of Eq. \ref{eq:sigusq}, \ref{eq:siguv} and \ref{eq:sigvsq_withsel}.  We measure the coefficients for 600 6dFGS mock catalogues, and the panels of the figure display their joint distribution (top row), and a histogram (bottom row).  The solid red lines indicate the predictions of the analytical model, which are in good agreement with the measurements.}
  \label{fig:6dfgs_coefficients}
\end{figure*}

As a convenient example to test these methods, we analyse the 600 mock catalogues for the 6-degree Field Galaxy Survey redshift sample in the redshift range $z < 0.1$, studied by \cite{2023MNRAS.518.2436T} and originally created by \cite{2018MNRAS.481.2371C}.   This selection function extends over the southern hemisphere for declinations $\delta < 0^\circ$, excluding a region around the Galactic Plane, and drops off steeply with increasing redshift owing to the magnitude-limited selection.  We ended the selection function in a cuboid of dimensions $600 \times 600 \times 300 \, h^{-1}$ Mpc.  We analysed each mock using the methods described in Sec.\ref{sec:model_error}, and compared the estimated values of the coefficients $\sigma_u$, $\sigma_v$ and $\sigma_{uv}$ from each realisation, with the analytical determinations based on Eq. \ref{eq:sigusq}, \ref{eq:siguv} and \ref{eq:sigvsq_withsel}. The results are displayed in Fig. \ref{fig:6dfgs_coefficients}, indicating that the analytical model provides a good description of the measurements.

% Don't change these lines
\bsp	% typesetting comment
\label{lastpage}
\end{document}